\documentstyle[prd,aps,epsf,preprint,citesort]{revtex}
\begin{document}
\draft
\tightenlines
\title{\hfill {\small UTEXAS-HEP-01-20} \\
\hfill  {\small MSUHEP-03161} \\ 
\hfill \\ {\large\bf Ultrahigh-Energy Neutrino-Nucleon Cross Sections \\ and
Perturbative Unitarity}}
\author{Duane A.~Dicus$^1$\footnote{PHDB057@utxvms.cc.utexas.edu}, Stefan Kretzer$^2$\footnote{kretzer@pa.msu.edu},
Wayne W.~Repko$^2$\footnote{repko@pa.msu.edu} and Carl
Schmidt$^2$\footnote{schmidt@pa.msu.edu} }
\address{$^1$Center for Particle Physics and Department of Physics,
University of Texas, Austin, Texas 78712 \\ $^2$Department of Physics and
Astronomy, Michigan State University, East Lansing, Michigan 48824}
\date{\today}
\maketitle 

\begin{abstract}\noindent
Unitarity relates the total cross section for neutrino-nucleon scattering
to the neutrino-nucleon forward scattering amplitude. Assuming the
validity of the perturbative expansion of the forward amplitude in the
{\em weak} coupling constant, we derive a unitarity bound on the inelastic
cross section. The inelastic cross section saturates this bound at a
typical neutrino energy $E_\nu \simeq 10^8\ {\rm GeV}$. This implies that
calculations of the inelastic cross section that use current parton
distribution functions and lowest order weak perturbation theory are
unreliable above this energy.
\end{abstract}

\baselineskip=18pt
\parskip=8pt

\section{Introduction}

The phenomenology of ultrahigh-energy (UHE) neutrinos and their detection
\cite{halzen} depends on the neutrino-nucleon total cross section, which has
been calculated in the standard model in
Refs.~\cite{fmr,gqrs96,gqrs98,gkr,kms}. A striking feature of all of these
predictions is the continued power-law-like growth of the cross sections with
$E_\nu$ at the highest energies. This rise with $E_\nu$ is directly related to
the very-low-$x$ behaviour of the nucleon parton distribution functions
(PDF's), which must be extrapolated below the regime of the current HERA data
\cite{hera}. It has even been argued \cite{JMPR,NS,GW,DKDBW} that cosmic ray
events will soon require a cross section even larger than that given by the
standard calculations.

On the other hand, unitarity relates the total scattering cross section to the
forward scattering amplitude. For neutrino-nucleon scattering, in contrast to
electron-nucleon scattering, the forward amplitude can be determined and used
to bound the total cross section.  This bound, based on lowest order
perturbation theory in the weak coupling, is independent of the PDF's and
implies that the inelastic cross section cannot rise indefinitely. In fact, it
cannot rise for more than about two decades in energy above the region now
covered by HERA.  Thus, either there will need to be a dramatic change in the
way parton distribution functions scale not too far above the HERA data, or
the assumption of weak perturbation theory is wrong.  In either case the
inelastic cross section will likely be very different at high $E_\nu$ than
what is predicted in the references above.

In Section 2 we discuss the relation between the cross section at high
$E_\nu$ and the low $x$ behavior of the distribution functions.  In
Section 3 we derive our bound and Section 4 contains conclusions.

\section{Ultrahigh-Energy Neutrino-Nucleon Cross
Sections}

At ultrahigh neutrino energies the total cross section is completely
dominated by its deep inelastic component. The deep inelastic scattering
(DIS) cross section can be evaluated within the QCD-improved parton model,
employing the universal parton distribution functions (PDF's) of the
nucleon \cite{cteq5,grv98,mrst} which are derived from fitting
photon-exchange dominated DIS measurements. Most important for high
energies are the recent HERA low-$x$ measurements \cite{hera}, because UHE
neutrino cross sections receive a dominant contribution from the
ultrasmall-$x$ region of DIS -- even below the range covered by HERA.

The total neutrino-nucleon cross section is given by
\begin{equation}
\label{intsig} \sigma_{\rm tot}^{\nu N \rightarrow X}(s = 2 M_N E_\nu )
=\int_0^{1} dx\ \int_{0}^{sx} dQ^2\ d^2\sigma^{\nu N \rightarrow X}/dx d
Q^2\,,
\end{equation}
with the standard (weak) DIS cross section of the form
\footnote{Sub-leading QCD corrections are implicitly included in the
evolution of $q(x)=q(x,Q^2),{\bar q}(x,Q^2)$; explicit
${\cal{O}}(\alpha_s^1)$ corrections to Eq.~(\ref{intsig}) are negligible
${\cal{O}}(1-2\%)$.}
\begin{equation}
\label{diffsig} \frac{d^2\sigma^{\nu N \rightarrow X}}{dx d Q^2} =
\frac{G_F^2}{\pi} \left(\frac{M_{W,Z}^2}{Q^2+M_{W,Z}^2}\right)^{2} \big[
q(x,Q^2) + {\bar q}(x,Q^2) (1-y)^2\big]\,,
\end{equation}
where $y = Q^2/(x s)$. The effective quark and anti-quark distributions
$q(x,Q^2)$ and ${\bar q}(x,Q^2)$ include the appropriate (electro-)weak
couplings for charged or neutral current DIS.  In what follows we will
always assume an isoscalar nucleus $[N=(p+n)/2]$. Note, that
Eq.~(\ref{intsig}) is well behaved only for neutrino DIS where -- contrary
to the case of photon exchange -- the heavy vector meson propagator in
Eq.\,(\ref{diffsig}) is non-singular
\footnote{ {\it i.e.}~imposing a ``$Q>$few GeV'' cut-off on the integral
in (\ref{intsig}) changes its value only marginally.} at $Q^2=xys=0$. Accordingly, the
following arguments are tied to the weak interaction framework, with no
analogue for the more familiar photon exchange process. For explicit
calculations we restrict ourselves to neutrino, rather than anti-neutrino,
scattering, with the understanding that sea-quark dominance in the UHE
limit does not discriminate between neutrinos and anti-neutrinos.

Recent evaluations of UHE cross sections have extrapolated PDF's below
$x<10^{-4}$ using power laws \cite{gqrs98}, radiative renormalization
group behaviour \cite{gkr} or BFKL-type resummations \cite{kms}. Within
the accuracy required for astrophysical phenomenology these approaches
agree and predict a steep rise of the low-$x$ parton density functions.
Any flattening of the PDF's at low $x$ due to gluon recombination is
naively expected to be absent at the UHE scale $Q\simeq M_W$ due to the
higher twist nature ${\cal{O}}({1/Q^2})$ of such recombination channels
\cite{recombination}.

To gain a feeling for the interplay between the UHE cross section and the
low-$x$ behaviour of the PDF's we can perform the $Q^2$ integral under the
approximation that we neglect the evolution of the PDF's. We
find\footnote{We have verified that the effects of evolution induce
corrections of only about 20\%.}
\begin{equation}
\sigma_{\rm tot}^{\nu N \rightarrow X}(s ) \simeq \frac{M_{W,Z}^2
G_F^2}{\pi} \int_{0}^1 d x
 \left[q(x)\left(\frac{{\hat s}}{1+{\hat s}}\right)  + {\bar q}(x)
\left(\frac{2}{{\hat s}}+1-2\left(\frac{1+{\hat s}}{{\hat s}^2}\right)
\ln(1+{\hat s})\right) \right]\ ,\label{IntOutQ}
\end{equation}
where ${\hat s}=sx/M_{W,Z}^2$ and the PDF's are evaluated at some fixed scale
({\it i.e.} $q(x)=q(x,M_{W,Z}^2)$, ${\bar q}(x)={\bar q}(x,M_{W,Z}^2)$). From
(\ref{IntOutQ}) it is straightforward to show that assumed behaviour of the
PDF's at $x\rightarrow 0$ determines the $E_\nu \rightarrow \infty$ asymptotic
behaviour of the total cross section \cite{mr}: A power-type $x\rightarrow 0$
behaviour $x q(x)\propto x^{-\beta}$ \cite{gqrs98} of the sea-quark PDF
translates into a power-type rise with energy $\sigma^{\nu N\rightarrow
X}_{\rm tot} \propto E_\nu^\beta$. Renormalization group evolution from a flat
low-scale input \cite{gkr} gives rise to a slower, but still significant,
growth with energy. This is easiest to see in the double-leading-logarithm
approximation \cite{dla}, in which the cross section can be seen analytically
\cite{mr} to rise less fast than any power but faster than a logarithm.
Although each of the different approximations of the small-$x$ behaviour gives
different quantitative predictions of the $E_\nu \rightarrow \infty$
asymptotics, we observe in Fig.\,\ref{bound} that they agree to a good
approximation up to energies as high as $E_\nu = 10^{12}\ {\rm GeV}$. In
particular, they all rise continuously with energy.

This rise, which is a direct consequence of the $x \rightarrow 0$ behaviour of
$q(x),\, {\bar q}(x)$, reflects mainly the HERA measurements of $F_2^{\gamma}\,
\simeq x\, \sum e_q^2 ( q + {\bar q} ) (x)$ at small-$x$. The question is,
therefore, up to what energy is the rise of $\sigma_{\rm tot}$ strictly
required by these data? The $ep$ collider HERA runs at a center of mass energy
of $\sqrt{s}\simeq 300\, {\rm GeV}$ corresponding to an equivalent neutrino
energy of $E_\nu^{\rm HERA} \simeq 5 \times 10^4\ {\rm GeV}$ \cite{heratot}.
However, leaving theoretical extrapolations aside, $\sigma_{\rm tot}$ for
$E_\nu > E_\nu^{\rm HERA}$ is not completely unknown because HERA probes the
{\it differential} $d \sigma / dx dQ^2$ in Eq.\,(\ref{diffsig}) down to much
lower $x$ (corresponding to higher $E_\nu$) than the average $\left< {\rm
log}_{10} x\right> \simeq -1.5$ in the {\it integrated } $\sigma_{\rm
tot}(E_\nu^{\rm HERA})$, albeit at a lower scale $Q < M_{W,Z}$ for the
$\gamma^\ast$-exchange process. Hence, the maximal $E_\nu$ up to which the
evaluation of $\sigma_{\rm tot}$ according to Eqs.\,(\ref{intsig}) and
(\ref{diffsig}) is fixed by HERA depends on the extent to which we trust the
NLO QCD scale-evolution of DIS structure functions. Recent estimates
\cite{nnlo} of NNLO corrections to parton evolution confirm a trustworthy
stability of the full singlet evolution at least down to $x\gtrsim 10^{-4}$
for $Q^2 \gtrsim 10\ {\rm GeV}^2$. This latter kinematical range is well
covered by HERA and the corresponding data are incorporated in the PDF sets of
Refs.\,\cite{cteq5,grv98,mrst}. Assuming nothing
more\footnote{I.e.~assumptions which are intrinsic to the perturbative
evaluation in Eq.~(\ref{intsig}).} than the universality of the PDF's and the
validity of the RGE within $10^{-4}\alt x,\, 10\ {\rm GeV}^2 < Q^2 < M_W^2$, we
can then consider $\sigma_{\rm tot}$ -- if evaluated according to
Eqs.\,(\ref{intsig}) and (\ref{diffsig}) -- as {\it effectively} covered by
HERA data up to neutrino energies where $\sigma_{\rm tot}$ becomes sensitive
to $x<10^{-4}$. This sets in\footnote{For illustration, see Fig.~3 in
\cite{gqrs96} or Fig.~3 in \cite{gkr}.} smoothly above $E_\nu \gtrsim 10^{6}\
{\rm GeV}$, which we indicate by the vertical line in Fig.~1. Note, this is a
very conservative estimate of the impact of the HERA data. Were we to trust the
PDF-mediated mapping of the HERA measurements onto the neutrino cross-section
down to $x\simeq 3 \times 10^{-5}$, and rely on the evolution from $Q^2=1.5\
{\rm GeV}^2$ to $M_W^2$, then the most recent HERA data \cite{h1new} would
cover $\sigma_{\rm tot}$ at $E_\nu \simeq 10^8\ {\rm GeV}$ except for a
correction of $\sim 25 \%$ from $x<3 \times 10^{-5}$. As we will deduce in the
next section, this would suggest HERA $\gamma$-exchange data are already
probing a perturbative unitarity bound in UHE neutrino scattering. However,
for now, we prefer not to speculate about shifting the vertical line in Fig.~1
to the right. A detailed statistical analysis, though, might do better and
provide stringent correlations between HERA data and $\sigma_{\rm tot}$ at
even higher $E_\nu $. Note, our limitation $x > 10^{-4}$ is also a safe
condition for the absence of recombination corrections \cite{recombination} at
HERA scales which are, accordingly, even more negligible at the UHE scale
$Q=M_{W,Z}$.

\section{Perturbative Unitarity}

Using unitarity of the $S$-matrix, we can relate the total $\nu N$ cross
section in (\ref{intsig}) to the imaginary part of the neutrino-nucleon
forward scattering amplitude:
\begin{equation}
\label{unitarity} \sqrt{ \lambda } \sigma_{\rm tot}^{\nu N \rightarrow
X}(s) = {\rm Im}\left[ T_{\{\nu N, \nu N\}} (s,t=0) \right]\,,
\end{equation}
where $\lambda =(s-M_N^2)^2$ and $s,t$ are the
standard Mandelstam variables. The elastic amplitude $T$ is related  to
the elastic cross section
\begin{equation}
\label{elastic} \frac{d \sigma_{\rm el}^{\nu N \rightarrow \nu N}}{d t} =
\frac{1}{16 \pi \lambda} \left|  T_{\{\nu N, \nu N\}} (s,t)\right|^2\,.
\end{equation}
Combining Eqs.~(\ref{unitarity}) and (\ref{elastic}) gives a general limit
on $\sigma_{\rm tot}$ in Eq.~(\ref{intsig})
\begin{equation}
\label{limit} \left. \frac{d \sigma_{\rm el}}{d t}\right|_{t=0} =
\frac{1}{16\pi} \left[ \frac{\left({\rm Re}[ T_{\{\nu N, \nu N\}}
(s,0)]\right)^2}{\lambda} + \sigma_{\rm tot}^2\right] \ge \frac{1}{16\pi}
\sigma_{\rm tot}^2\ \ \ .
\end{equation}
We note that the inequality Eq.\,(\ref{limit}), as derived, holds strictly for
each spin and isospin state of the nucleon; however, it is straightforward
to show that\footnote{The difference between the left and right hand sides
of Eq.~(\ref{spins}) is the variance of elementary probability theory and
therefore positive.}
\begin{equation}
\frac{1}{4}\sum \sigma^2_{\rm tot} \ge \left[ \frac{1}{4}\sum \sigma_{\rm
tot}\right]^2\ ,\label{spins}
\end{equation}
where the sum is over spin and isospin of the nucleon. Thus, we can just
as well use the spin/isospin averaged cross sections in
Eq.~(\ref{limit})\footnote{In fact, due to constraints on spin and isospin
asymmetries in DIS  scattering, one would expect the
difference between the right-hand and left-hand sides of Eq.~(\ref{spins})
to be negligible in the high energy limit.}.

The inequality in Eq.\,(\ref{limit}) is a standard statement of the optical
theorem and can be found in many textbooks.  It follows strictly from the
positivity of $({\rm Re}T)^2$ and does not rely on any perturbative
expansion.  We now consider its implications in the context of a
perturbative calculation in the weak coupling.  Expanding the elastic
amplitude to lowest power in the weak coupling $g$ and using the most
general $Z$-nucleon coupling (see Fig.\,\ref{secondorder}), we obtain
\begin{eqnarray}
T&=& \left( \frac{g}{4 \cos \theta_W}\right)^2 \ {\bar u}_{\nu}(k^\prime
)\, \gamma^\alpha\, (1 - \gamma_5)\, u_{\nu}(k) \
\frac{1}{t-M_Z^2}\nonumber\\ \label{loword}
&& \times\ {\bar u}_{N}(p^\prime )\, \left[\gamma_\alpha\, f_1(q^2 )
 + i\sigma_{\alpha \beta}\, q^\beta\, f_2(q^2 ) +
q_\alpha\, f_3(q^2 )+ \gamma_\alpha\,\gamma_5\, g_1(q^2 )\right.\\[4pt]
&& \left.+ i \sigma_{\alpha \beta}\, \gamma_5\, q^\beta\, g_2( q^2
) +  q_\alpha\, \gamma_5\, g_3(q^2 ) \right]\,
 u_{N}(p) +{\cal{O}}(g^4)\,.\nonumber
\end{eqnarray}

In the forward direction only the form factors $f_1(0)$ and $g_1(0)$
contribute. Using the standard model $Z$-boson current and isospin symmetry,
the values of the form factors at $q^2=0$ can be expressed in terms of the
nucleon isospin operator $T_3$ and charge operator $Q$ as
$f_1(0)=(2\,T_3-4\sin^2\theta_W\,Q)$ and $g_1(0)=-2\,T_3\,g_A$, where
$g_A=1.27$ is the measured value \cite{pdg} of the axial vector coupling
constant. For an isoscalar nucleus we obtain
\begin{equation}
\label{sigel} \left. \frac{d \sigma_{\rm el}^{(2)}}{d t}\right|_{t=0} =
\frac{G_F^2}{8 \pi} \left[\frac{1+(1-4\sin^2\theta_W)^2}{2}+g_A^2
 \right] \,.
\end{equation}
Accordingly, the inequality Eq.\,(\ref{limit}) gives
\begin{equation}
\label{value} \sigma_{\rm tot} \lesssim 9.3 \times 10^{-33}\ {\rm cm}^2\,.
\end{equation}
The value on the right-hand side of Eq.\,(\ref{value}) is shown as a horizontal
line in Fig.\,\ref{bound}. We observe that the inequality is violated for
$E_\nu \gtrsim 2\times10^8\ {\rm GeV}$. Note that the precise value of the
bound is rather insensitive to the values of $f_1(0)$ and $g_1(0)$ because
it depends on the square root of the factor in the square brackets of
Eq.\,(\ref{sigel}).

HERA information on PDF's determines $\sigma_{\rm tot}$ for neutrino
energies as high as $E_\nu \gtrsim 10^6\ {\rm GeV}$. Perhaps large changes
in the PDF's will set in within the range of $x$ appropriate for $E_\nu
\sim10^6-10^8\ {\rm GeV}$ such that the bound is respected for larger
$E_\nu$. These corrections would presumably arise as a softening of the
gluon and sea quark distributions at small $x$.

Alternatively, one can note that the bound Eq.\,(\ref{value}) is obtained by
comparing an ${\cal O}(g^4)$ expression for $\sigma_{\rm tot}$ with an
${\cal{O}}(g^2)$ expression for the forward elastic scattering amplitude.
Since unitarity is an order-by-order statement within perturbation theory,
to be strictly rigorous we must also include the absorptive ${\cal
O}(g^4)$ contribution to the forward scattering amplitude.  In this way,
unitarity in the ultrahigh energy limit can in principle be restored, but
at the cost of large ${\cal{O}}(g^4)$ corrections to the forward
amplitude.

For illustration,  we write the elastic amplitude symbolically as
\begin{equation}
\label{nextord}
T = g^2 T^{(2)} + g^4 T^{(4)} + \ldots\,.
\end{equation}
where we have calculated the $T^{(2)}$ part above and  the $T^{(4)}$ term
results from the exchange of a pair of virtual $Z^0$'s or $W^\pm$'s in the
elastic scattering, as illustrated in Fig.\,\ref{fourthorder}. Using this
notation
\begin{equation}
\label{diffpert} \left. \frac{d \sigma_{\rm el}}{d t}\right|_{t=0}
\propto\ g^4 {T^{(2)}}^2 + 2g^6 T^{(2)}\ {\rm Re}[T^{(4)}] + g^8 \left[
\left( {\rm Re}[T^{(4)}]\right)^2 + \left( {\rm Im}
[T^{(4)}]\right)^2 + \ldots\right]+ \ldots\,,
\end{equation}
and
\begin{equation}
\label{totpert} \left(\sigma_{\rm tot}\right)^2 \propto g^8 \left( {\rm
Im}[T^{(4)}]\right)^2+ \ldots\,.
\end{equation}
From these equations we see that the unitarity condition Eq.\,(\ref{limit})
can be satisfied; however it requires that the {\it higher order} $g^4 T^{(4)}$
term become larger than the {\it lower order} $g^2 T^{(2)}$ term in the high
energy limit. So, in this approach, calculating to lowest order in the weak
coupling $g$ is not a good approximation for the elastic differential cross
section at high energy.  But if the perturbation expansion is unreliable in
Eq.\,(\ref{diffpert}), then it is also reasonable to expect higher order
corrections to Im$\,T$ to be larger than the term shown, both in
Eq.\,(\ref{diffpert}) and in Eq.\,(\ref{totpert}). Since this is the term on
which all calculations have been based it implies that these calculations may
not be reliable at very high energies. If the perturbation expansion doesn't
work in Eq.\,(\ref{diffpert}) then we must question its validity in
Eq.\,(\ref{totpert}).

\section{Conclusions}

\begin{figure}[h]
\hspace{1.0in}\epsfysize=4in \epsfbox{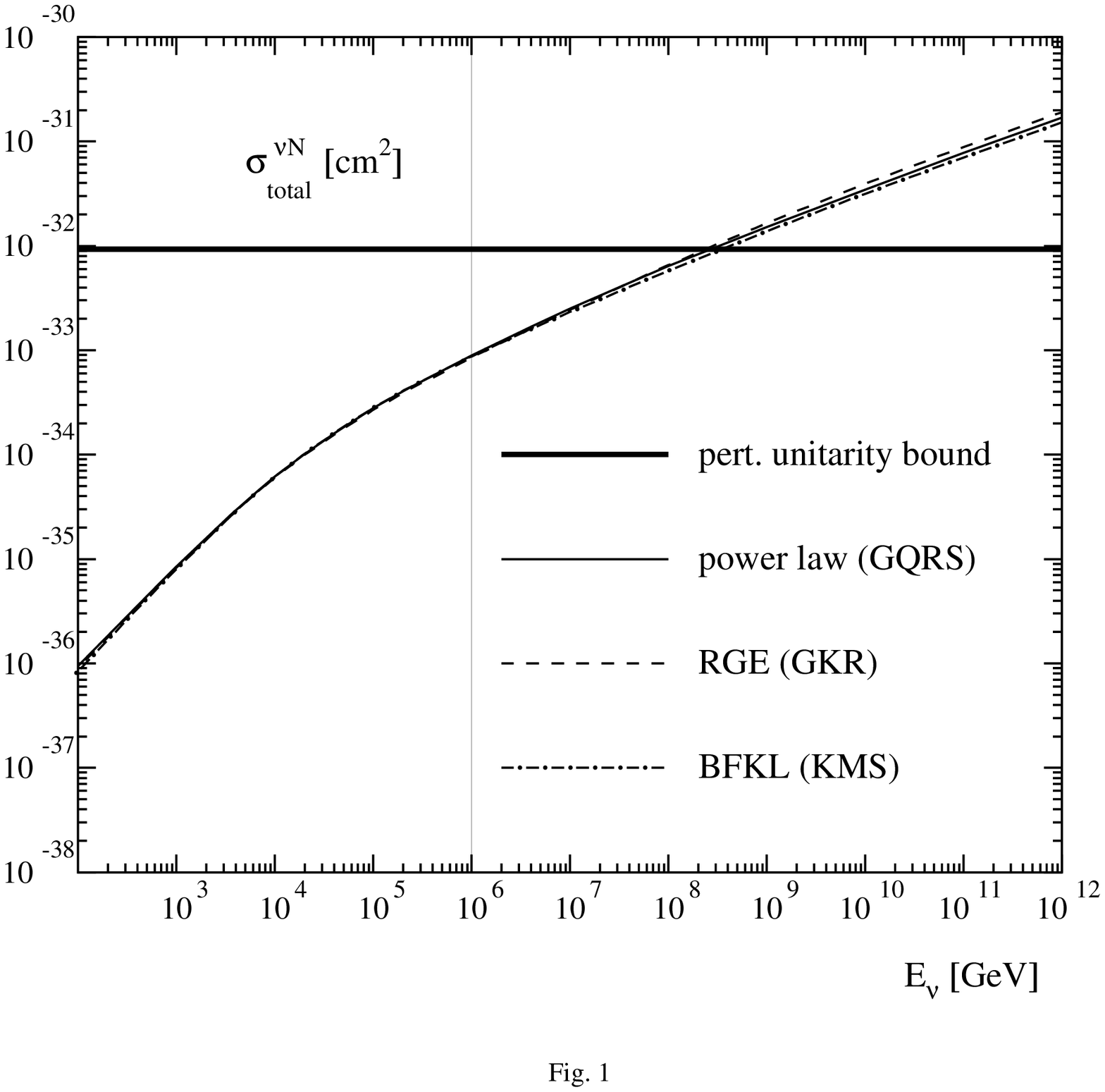}
\caption{\footnotesize Perturbatively evaluated QCD neutrino-nucleon total
cross sections from a power law (GQRS)\protect\cite{gqrs98}, renormalization
group evolution (GKR)\protect\cite{gkr}, or BFKL (KMS)\protect\cite{kms}
approach towards ultrasmall-$x$ structure functions. The thick horizontal line
is the lowest order perturbative unitarity bound in Eq.~(\ref{value}). The
vertical line is a conservative upper bound on the $E_\nu$-range effectively
covered by HERA, {\it i.e.}~the perturbative cross section for energies $E_\nu
\protect\alt 10^6\ {\rm GeV}$ receives no significant contribution from the $x
\protect\alt 10^{-4}$ regime of DIS.\label{bound} }
\end{figure}

\begin{figure}[h]
\hspace{.5in}\epsfysize=2in \epsfbox{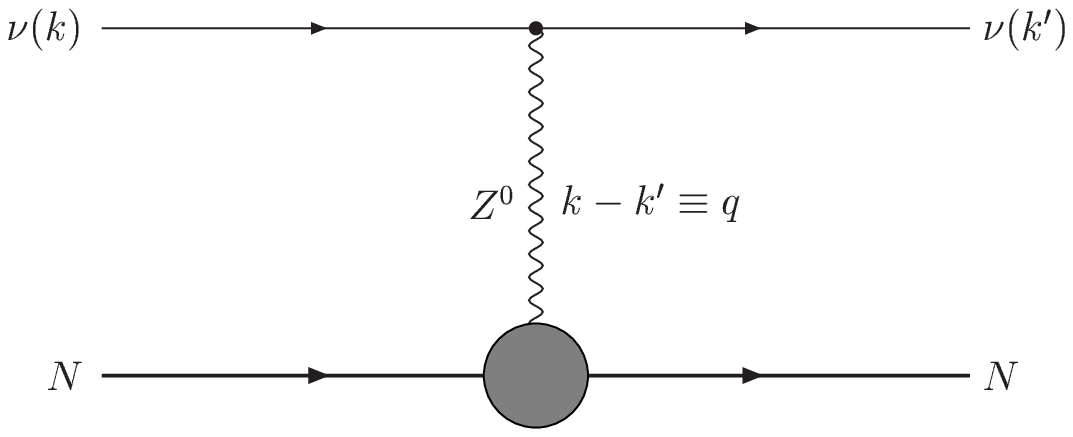}
\vspace{10pt}
\caption{Lowest order [${\cal{O}}(g^2)$] elastic amplitude $T_{\{\nu N,
\nu N\}}$ corresponding to Eq.~(\ref{loword}).\label{secondorder}}
\end{figure}

\begin{figure}[h]
\hspace{.5in}\epsfysize=2in \epsfbox{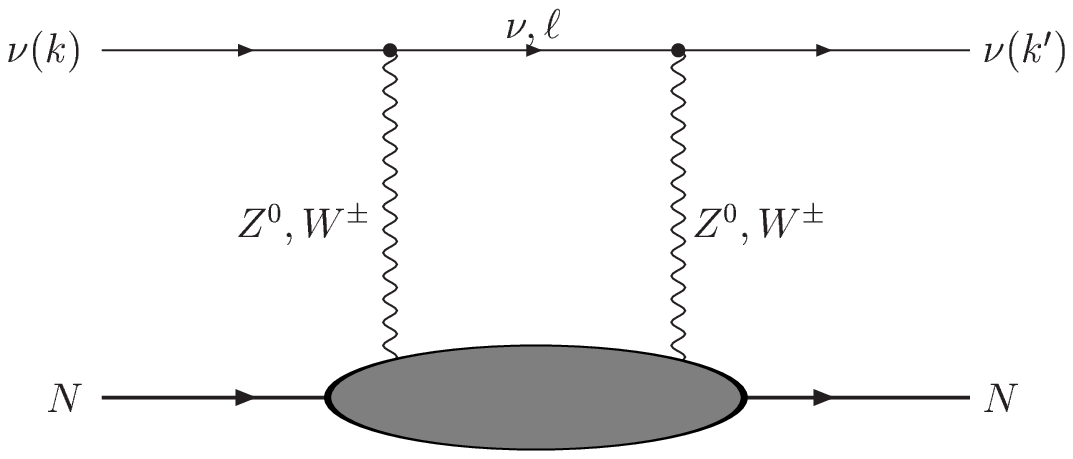}
\vspace{10pt}
\caption{Next order [${\cal{O}}(g^4)$] elastic Amplitude $T_{\{\nu N, \nu
N\}}$ corresponding to $T^{(4)}$ in Eq.~(\ref{nextord}).\label{fourthorder}}
\end{figure}

We compared perturbative predictions for neutrino-nucleon total cross
sections at ultrahigh neutrino energies to a unitarity bound derived from
the corresponding elastic neutrino scattering amplitude in the forward
direction. At the non-perturbative level the bound is absolutely
stringent. A lowest order expansion in powers of the weak coupling leads
to a bound which is saturated at an energy surprisingly close to what is
effectively covered by HERA measurements. Thus, for the largest energies
relevant to neutrino astronomy, sizable PDF corrections could reside in
$\sigma_{\rm tot}$. Alternatively, the large apparent ${\cal{O}}(g^4)$
total cross section should manifest itself through the corresponding term
in the elastic amplitude as well. But if the perturbation expansion is not
valid for the elastic amplitude how can we trust it for the inelastic
cross section?  In this case also $\sigma_{\rm tot}$ could be very
different for large $E_\nu$ than the curves shown in Fig.\,\ref{bound}.

We have ignored the possibility that unitarity is satisfied by a new
interaction -- physics beyond the standard model \cite{JMPR,NS,GW,DKDBW}. But
even in this alternative our basic conclusions remain the same: something
dramatic must happen at energies close to the current energy and the existing
calculations of $\sigma_{\rm tot}$ should not be trusted at large neutrino
energy.

If the bound is correct, and the PDF's change at low $x$, then $\sigma_{\rm
tot}$ will be smaller at high $E_\nu$ than what is shown in Fig.\,\ref{bound}.
If perturbation theory has broken down then $\sigma_{\rm tot}$ could be
anything - perhaps even large enough to explain the cosmic ray data without
the need for new physics. Or perhaps the values shown in Fig.\,\ref{bound}
could turn out to be correct; maybe weak perturbation theory only breaks down
for the elastic cross section. Without a better knowledge of the PDF's at low
$x$, there is no way to know.

\acknowledgements It is a pleasure to thank Karol Lang, Wu-Ki Tung and Scott
Willenbrock for helpful conversations and G. Domokos, M. Gl\"uck, and S.
Kovesi-Domokos for their comments. This research was supported in part by the
National Science Foundation under Grant PHY-0070443 and by the United States
Department of Energy under Contract No. DE-FG03-93ER40757.

\end{document}